\documentclass[twocolumn,english,ruled]{IEEEtran}
\usepackage[latin9]{inputenc}
\usepackage{babel}
\usepackage{booktabs}
\usepackage{algorithm2e}
\usepackage{amsmath}
\usepackage{amsthm}
\usepackage{stackrel}
\usepackage{graphicx}
\usepackage[unicode=true,
 bookmarks=true,bookmarksnumbered=true,bookmarksopen=true,bookmarksopenlevel=1,
 breaklinks=false,pdfborder={0 0 0},pdfborderstyle={},backref=false,colorlinks=false]
 {hyperref}
\hypersetup{pdftitle={Your Title},
 pdfauthor={Your Name},
 pdfborderstyle={},pdfborderstyle={},pdfborderstyle={},pdfborderstyle={},pdfborderstyle={},pdfborderstyle={},pdfborderstyle={},pdfpagelayout=OneColumn,pdfnewwindow=true,pdfstartview=XYZ,plainpages=false}
\usepackage{breakurl}

\makeatletter

\providecommand{\tabularnewline}{\\}

\theoremstyle{plain}
\newtheorem{thm}{\protect\theoremname}
\theoremstyle{definition}
\newtheorem{defn}[thm]{\protect\definitionname}
\theoremstyle{plain}

\theoremstyle{plain}
\newtheorem{lem}[thm]{\protect\lemmaname}

\usepackage{babel}
\usepackage{babel}
\usepackage{babel}
\usepackage{babel}
\usepackage{babel}
\usepackage{babel}
\usepackage{babel}
\usepackage{babel}

\usepackage{pifont}

\usepackage{booktabs}
\usepackage{multirow}
\usepackage{makecell}
\usepackage{threeparttable}
\usepackage{multicol}
\usepackage{algorithmic}
\ifCLASSOPTIONcompsoc
\else
\fi

\providecommand{\definitionname}{defn}
\providecommand{\theoremname}{thm}

\providecommand{\definitionname}{defn}
\providecommand{\lemmaname}{lem}
\providecommand{\theoremname}{thm}

\providecommand{\definitionname}{Definition}
\providecommand{\lemmaname}{Lemma}
\providecommand{\theoremname}{Theorem}

\providecommand{\definitionname}{Definition}
\providecommand{\lemmaname}{Lemma}
\providecommand{\theoremname}{Theorem}

\providecommand{\definitionname}{Definition}
\providecommand{\lemmaname}{Lemma}
\providecommand{\theoremname}{Theorem}

\providecommand{\definitionname}{Definition}
\providecommand{\lemmaname}{Lemma}
\providecommand{\theoremname}{Theorem}

\providecommand{\definitionname}{Definition}
\providecommand{\lemmaname}{Lemma}
\providecommand{\theoremname}{Theorem}

\providecommand{\definitionname}{Definition}

\providecommand{\lemmaname}{Lemma}
\providecommand{\theoremname}{Theorem}

\makeatother

\providecommand{\corollaryname}{Corollary}
\providecommand{\definitionname}{Definition}
\providecommand{\lemmaname}{Lemma}
\providecommand{\theoremname}{Theorem}

\begin{document}

\title{Chargeable Sweep Coverage Problem}

\author{{\normalsize{}Dieyan Liang$^{a}$ and Hong Shen$^{a}$}\\
\textit{\normalsize{}$^{\text{a}}$School of Computer Science and
Engineering, Sun Yat-Sen University, China}}
\maketitle
\begin{abstract}

Sweep coverage is to realize the periodic coverage of targets by planning the periodic sweeping paths of mobile sensors. It is difficult for sensors to reduce energy consumption by reducing the moving distances. Therefore, charging technology is the best way to extend the lifetime of the sweep coverage network. This paper studies the sweep coverage of rechargeable sensors: the sensors are rechargeable, constantly sweep between the target points and the charging stations not only tp meet the periodic coverage requirement of the target points, but also need to return to the charging stations during the charging period to avoid running out of energy. This paper proposes the general definition of Chargeable Sweep Coverage (CSC) problem for the first time, and studies the complexity of the CSC problem by analyzing CSC problems under different constraints, and then proposes two kinds of CSC problems under special constraints: 1) The sensors need to return to their original charging stations for charging; 2) The sensors can go to  different charging stations for charging, and the number of charging stations is 2. Both of these problems are NP-hard. In this paper, these two problems are modeled as the maximum set coverage problem, and the approximation algorithms are obtained by reducing the number of candidate paths to polynomials. The validity and scalability of the proposed algorithms is proved by theoretical proof and experimental simulation. 

\end{abstract}


\section{Introduction}

Based on the wide application of mobile sensors in wireless sensor networks (WSNs), sweep coverage has become a hot issue in current research. The sweep coverage problem can be applied to scenarios where the targets do not need long-term coverage, but only need periodic coverage, such as the periodic patrol of the security system, the periodic collection of information, and so on. In addition, the models for the sweep coverage problem can also be used in the applications in the emerging mobile crowd sensing\cite{wu2019task,huang2018efficient}. Considering that the energy consumption of moving is far greater than the energy consumption of sensing and communication,  mobile sensors would consume a lot of energy in sweep coverage. And sensors have limited energy, which leads to the lifetime of the sweep coverage network shorter. How to extend the lifetime of sweep coverage network is an important issue. In recent years,  the charging technology has gradually matured and the cost has been reduced,  so wireless charging and solar charging technologies have been gradually introduced into WSNs. Solar charging is generally used as a supplementary energy due to unstability affected by the weather and location. And wireless charging has the risk of electromagnetic exposure that leads to a small coverage area and requires short-distance charging. However, wireless charging is very suitable for sweep coverage networks, because the mobile sensors can move to the charging stations for short-distance charging. At present, there are only few studies considering the charging  in the sweep coverage problem\cite{gorain2019approximation,gao2020cooperative}.


This paper mainly studies the  Chargeable Sweep Coverage (CSC) problem. Each sensor travels periodically on the target area. It not only needs to meet the periodic coverage requirements of the target point, but also needs to meet its charging requirements. That is,  each sensor needs to pass through at least one charging station in evergy charing period $T_c$, and at the same time makes the target covered by at least one sensor in every sweep period $T_t$. This charging model can be used in all application scenarios of sweep coverage, and greatly improves the autonomy of the sensor network and its network lifetime.

This paper first proposes the definition of the chargeable sweep coverage (CSC) problem and two CSC problem under  special constraints. These two special constraints are:  when the sensors must return to the original charging stations for charging, and when the sensors can go to any charging stations for charging but the number of charging stations is 2. We prove that these two CSC problem under special constraints are is Non-deterministic Polynomial hard(NP-hard). Then, two approximation algorithms for the two CSC under special constraints are given. The main contributions are as follows:
\begin{enumerate}
	\item We put forward the general chargeable sweep coverage problem and the chargeable sweep coverage problem under two special constraints, which prove that these three problems are NP-hard.
	\item Under the constraint that the sensors must return to the original charging stations to charge, when $T_c\ge T_t$,  an approximation algorithm with an approximation ratio of about $\frac{2}{5}$ is proposed; when $T_t>T_c $, the approximation ratio is related to $\hat{Q}=T_t/T_c$, then an approximation algorithm with the approximation ratio interval $[\frac{1}{6},\frac{1}{3}]$ is proposed.
	\item Under the constraint that the sensors can go to any charging stations to charge and the number of charging stations is 2, when $T_c\ge T_t$, an approximation algorithm with an approximation ratio of about $\frac{1}{4}$ is proposed; when $T_t >T_c$, the approximation ratio is related to $\hat{Q}=T_t/T_c$, an approximation algorithm with the approximation ratio interval $[\frac{1}{6},\frac{3}{10}]$ is proposed.
\end{enumerate}

The rest of this paper is organized as follows. Section \ref{sec:CSC-related-works}gives the related works. Section \ref{sec:CSC-model} gives the system model of the CSC problem, the problem description, and studies the difficulty of the CSC problem by the related work studied. Two CSC problems under special constrains are also proposed. Section \ref{sec:CSC-RCSC} gives an approximation algorithm for the CSC problem when the sensors must return to the original charging stations. Section \ref{sec:CSC-2CSC} gives an approximation algorithm for the CSC problem when the sensors can go to any charging stations to charge but the number of charging stations is 2. Section \ref{sec:CSC-experiment} performs simulation experiments. Section \ref{sec:CSC-conclution} is the conclusion of this paper.

\section{Related works}
\label{sec:CSC-related-works} 

There are three optimization objectives for sweep coverage problem, including maximum coverage quality, minimum number of sensors, and minimum sweep period. Most of recent works are focus on the minimum number of sensors and the minimum sweep period.

The minimum sensor sweep coverage (MSSC) problem to find the minimum number of sensors to periodically cover all the target points distributed on the plane is first proposed by Cheng et al\cite{cheng2008sweep}, in which the velocities of sensors are the same, and so are the sweep periods of targets. The problem was proved to be NP-hard through reducing the travelling salesmen problem (TSP) to it and it is proved  there did not exist effective local algorithms or  an approximation algorithm with an approximation ratio 2. 
Gorain et al. \cite{gorain2015approximation} then proposed 3-approximation algorithms to solve the MSSC problem. So far, this is still the approximation algorithm for the MSSC problem with the bestl approximation ratio. It is proved that when the velocities are different, the approximation algorithm with constant approximation ratio does not exist. When the target points have different sweep periods, An approximation algorithm with an approximation ratio of $O(log\rho)$ was proposed for the MSSC problem, where $\rho$ is the ratio of the maximum sweep period to the minimum sweep period of the targets. The area sweep coverage problem and the fence sweep coverage problem are also NP-hard. The area sweep coverage problem has an approximation algorithm with an approximation ratio of $(\sqrt{2}+\frac{2-\sqrt{2}}{mn})$\cite{gorain2014approximation}, the fence sweep coverage problem has a 2-approximation algorithm\cite{gorain2014line}.

 Pasqualetti et al. \cite{pasqualetti2012cooperative} began to study the minimum period sweep coverage (MPSC) problem. The scenarios include curves, trees, and cyclic graphs. When the target points are on the curve, the optimal algorithm with a time complexity of $O(nlog(\epsilon^{-1}))$ can be obtained; when the target points are on the acyclic graph (tree) and the number of sensors is constant, the polynomial-time optimal algorithm can be obtained; when the target points are on the cyclic graph, an 8-approximation algorithm can be obtained. Collins et al. \cite{collins2013optimal} extended the MPSC problem to segment fences. Gao et al.  studied the MPSC problem when the target points are distributed on a two-dimensional plane, and proposed a 5-approximation algorithm for the case when sensors were scheduled by separation strategy\cite{gao2018approximation}, and 4-approximation algorithm for the case when the sensors were under cooperative strategy\cite{gao2020cooperative}.  The MPSC problem where the  target is a graph is also stuedied\cite{gao2018approximation}, that is, the target includes the vertices of the graph and the edges between the vertices. Different from the paper\cite{gao2018approximation}, the paper \cite{pasqualetti2012cooperative} needs to periodically cover the vertices in the graph, while the edges in the graph (trees and cyclic graphs) only constrain the movement path of sensors. 

Besides, Zhao et al. \cite{zhao2012mobile, zhang2019timely} considered the constraint of data transmission delay, and proposed two heuristic algorithms to solve the dual delay constrained minimum velocity problem (DDC-MVS) when the data transmission delay and the sweep period are inconsistent. Chen et al. \cite{chen2016efficient} shortened the movement trajectory of the sensor considering the influence of the coverage radius of the sensor. Yu et al. \cite{yu2017participant} considered the robustness of coverage and proposed the $k$-sweep coverage problem. 

One of major challenges of wireless sensor networks is because of their limited energy.  Gorain et al. \cite{gorain2016solving} considered the difference in energy consumption between static sensors and dynamic sensors, and proposed two questions, one is the sweep coverage problem of minimizing energy consumption, and the other is the sweep coverage problem of the minimum sensor with limited energy, and give the 2-approximation algorithm and the $5+\frac{2}{\alpha}$-approximation algorithm respectively. In recent years, due to the development of charging technology, sensors with rechargeable battery  have also begun to be widely used in various fields. There are currently two main charging methods, one is solar-powered\cite{shi2016adaptive, shi2017constructing, shi2018coverage, wang2016hybrid}, and the other is mobile charging\cite{wang2016hybrid,gorain2019approximation,gao2020cooperative}. The paper \cite{wang2016hybrid} measures the pros and cons of the two charging methods, and proposes a hybrid framework to achieve the coverage quality and energy balance. The paper \cite{shi2016adaptive, shi2018coverage} considers the imbalance of sensor energy consumption, adjusts the sleep strategy, and uses solar charging to achieve coverage quality and extend network timelife. The paper \cite{shi2017constructing} considers the directionality of wireless charging, and minimizes the energy consumption of the charging nodes while statisfying coverage quality.

Due to the continuous movement of the mobile sensors in the sweep coverage network, mobile charging technology is more suitable to use to solve the  challenge of limited energy. Gorain et al. \cite {gorain2019approximation} first introduced a mobile sensor with charging capability into the sweep coverage problem, and proposed the energy-constrained fence sweep coverage problem. The mobile sensor needs to return to the base station for charging every charging period $T$, and realize the $t$-sweep coverage of the fence, $t\neq T$. Although it is impossible to prove its difficulty, the $\frac{13}{3}$-approximation algorithm is proposed. Gao et al. \cite{gao2020cooperative} takes into account the limitation of data storage capacity and energy, but does not consider battery capacity. It only requires that each sensor must pass through a base station on the sweeping path for data clearing and energy supplementation. A 6-approximation algorithm was proposed for finding the minimum sweep period under this constraint.

\section{Network model and problem description}
\label{sec:CSC-model}

\subsection{Network Model}

The network model of the CSC problem is showed below. Assuming that the charging stations and targets are distributed in the metric space, the network model can be represented by an undirected complete graph $G(\mathcal{T}\cup\mathcal{C}, \mathcal{E})$, where $\mathcal {T}=\{t_{1},t_{2},..,t_{N}\}$ represents the set of targets, $\mathcal{C}=\{c_{1},c_{2}, ..,c_{K}\}$ represents $K$ charging stations, $d: \mathcal{E}\rightarrow R^{+}$ is the weight of the edge $e\in\mathcal{E}$, which conforms to triangular inequality constraint, that is, for any three points $i, j, k\in \mathcal{T}\cup\mathcal{C}$, all satisfy $d(i,j)=d(j,i)$ and $d(i,j)\le d(i,k)+d(k,j)$.
Given $M$ sensors $\mathcal{S}=\{s_{1},s_{2},..,s_{M}\}$ with rechargeable batteries, each of them can sweep at the velocity $V$ for $T_c$ time when the battery is fully charged. be charged. Each target in $\mathcal{T}$ needs to be covered by sensor every $T_t$ time, referred to as "$T_t$-sweep covered" or "covered" for simplication. Since the sensors' moving distances are much larger than the sensing radius of the  sensors, this study simplified the coverage model, thinking that the sensor cover the target  only when it moves to the position of the target. Also, for simplication, the analysis in this paper assumes that the charging time $T_r$ can be ignored.
It is also assumed that the relationship between $T_c$ and $T_t$ is an integer multiple, that is, when $T_c\ge T_t$, then $T_c/T_t=Q$, where $Q\ge1$ is a positive integer; when $T_c< T_t $, then $T_t/T_c=\hat{Q}$, where $\hat{Q}\ge2$ is also a positive integer.

\subsection{The description and the Hardness of CSC problem}
\label{subset:description}

Under this network model, the definition of the CSC problem is as follows:

\begin{defn}
	\textbf{chargeable sweep coverage problem}: Given a set of sensors $\mathcal{S}=\{s_{1},s_{2},..,s_{M}\}$ with the same velocity $V$, a set of charging stations $\mathcal{C}=\{c_{1},c_{2},..,c_{K}\}$ and a set of target points $\mathcal{T}=\{ t_{1},t_{2},..,t_{N}\}$  distributed on a plane, each sensor needs to go to any charging station $c\in\mathcal{C}$ every $T_c$ time to charge.  $N_c\le N$ is a positive integer, ask whether the given sensors can $T_t$ sweep cover $N_c$ target points.
\end{defn}

Next, we study the CSC problem under several special situations. These CSC problems under some special situations can be transformed into some problems that have been studied, including the orienteering problem, the  team orienteering problem, and the distance constrained vehicle routing (DCVR) problem. These problems are all NP-hard problems. We first give the definitions below.

\begin{defn}
	\label{def:CSC-DVRP}
	\textbf{Distance constrained vehicle routing problem}\cite{laporte1984two}: Given a set of vertices in a metric space, a specified depot, and a distance bound $L$, find a minimum cardinality set of tours originating at the depot that covers all vertices, such that each tour has length at most $L$.
\end{defn}

\begin{defn}
	\label{def:CSC-OP}
	\textbf{the orienteering problem}\cite{zhao2012mobile}: 
	given an edge-weighted graph $G=(V,E)$, two nodes $s, t\in V$ and a length limit $L$, find an $s-t$ walk in $G$ of total length at most $L$ that maximizes the number of distinct nodes vistited by the walk.
\end{defn}

\begin{defn}
	\label{def:CSC-TOP}
	\textbf{Team Orienteering Problem}\cite{xu2020approximation}:
	To find $M\ge1$ paths of length at most $L$, which starts at node $s$ and ends at node $t$, such that the profit sum of serving the nodes in the $K$ paths is maximized.
\end{defn}

When $s=t$, The orienteering problem is also called the rooted orienteering problem, The team orienteering problem is called the rooted team orienteering problem .

It can be seen from the above definition that when the number of charging stations $K=1$, the number of sensors $M=1$, and $T_c/T_t=1$, the CSC problem can be transformed into the rooted orienteering problem. When the number of charging stations $K=1$, the number of sensors $M\ge 1$, and $T_t/T_c=\hat{Q}$, the CSC problem can be  transformed to the rooted team orienteering problem with the number of loops $M\times \hat{Q}$. When the number of charging stations $K=1$, to find the minimum number of sensors covering all targets, then the CSC problem can be transformed into the DCVR problem. The table \ref{tab:CSC-related-works} lists the best approximation algorithms for these problems.

\begin{table*}[htbp]
	\caption{Existing research results}
	\centering{}%
	\begin{threeparttable}
		\begin{tabular}{|c|c|c|c|c|c|c|}
			\hline
			Target & \makecell[c]{Number of \\ Charging Piles} & \makecell[c]{Number of \\Sensors} & Covering Target & T\_c/T\_t & approximation ratio \tabularnewline
			\hline
			\multirow{3}{*}{\makecell[c]{max\\target\\cover}} & 1 & 1 & vertices of Euler graph & 1 & $\frac{1}{1+\epsilon}$ \cite{chen2006orienteering} \tabularnewline
			\cline{2-6}
			& 1 & 1 & vertices of graph & 1 & 1/2\cite{paul2020budgeted} \tabularnewline
			\cline{2-6}
			& 1 & M & verticesof graph & 1/Q & $(1-(1/e)^{\frac{1}{2}})$\cite{xu2020approximation} \tabularnewline
			\hline
			\multirow{2}{*}{\makecell[c]{Minimum \\sensor\\number}} & 1 &-& vertices of graph & 1 & $(O(log(1/\epsilon)),1+ 1/\epsilon)$\cite{nagarajan2012approximation}* \tabularnewline
			\cline{2-6}
			& 1 &-& Barrier &-& \makecell[c]{$\frac{13}{3}$(whether NP is \\difficult or not is unknown)}\cite{gorain2019approximation}\tabularnewline
			\hline
		\end{tabular}
		\begin{tablenotes}
			\footnotesize
			\item[*]The paper \cite{nagarajan2012approximation} provides a $(O(log\frac{1}{\epsilon}),1+\epsilon)$-bicriteria approximation algorithm for DVRP problems, which means, for any $\epsilon>0$, if the limit distance $D$ can be relaxed to $(1+\epsilon)D$, then the algorithm can obtain a $O(log\frac{1 }{\epsilon})$ approximation value. The inapproximation ratio of the DVRP problem  is $\frac{3}{2}$.
		\end{tablenotes}
	\end{threeparttable}
	\label{tab:CSC-related-works}
\end{table*}

Form the table \ref{tab:CSC-related-works}, it is known that the CSC problem is so a trivial problem. In this paper, we consider the CSC problem under two special constraints. One is that each sensor must return to its original charging station for charging. This situation occurs when the charging stations are far away, that is, the length of the shortest path between each charging station is $SP(c_i,c_j)=d(c_i,c_j)>L_c$, where $c_i,c_j\in \mathcal{C}$, and $i\neq j$; or when the sensors and charging stations have matching models. We define the CSC problem in this case as a Restricted Chargeable Sweep Coverage (RCSC) problem. The other is that the sensors can go to any charging stations for charging, but the number of charging stations is two. We define this problem as a 2-chargeable sweep coverage (2-CSC) problem. In the RCSC problem, when the number of charging stations $K=1$, the number of sensors $M=1$, and the charging period and sweep period are the same, i.e,  $L_c=L_t$, the RCSC problem can be transformed into a rooted orienteering problem. It is NP-hard. In the 2-CSC problem, if the distance between the charging stations is far, greater than $VT_c$, the number of sensors $M=1$, and the charging period and sweep period are the same, i.e., $L_c=L_t$, the 2-CSC problem can be transformed into the problem of choosing the larger one between two rooted orienteering problems. It is also NP-hard. Therefore, we get the theorem \ref{thm:CSC-NPH}.

\begin{thm}
	\label{thm:CSC-NPH}
	Both the RCSC problem and the 2-CSC problem are NP-hard.
\end{thm}

For the convenience of description, we call a loop that starts from any charging station $c\in\mathcal{C}$ with a length of $L$ and returns to $c$ without passing through other charging stations as $c$-loop $r_c(L)$. A loop starts from any charging station $c_i\in\mathcal{C}$ and  pass at least one other charging station $c_j\in\mathcal{C}$ ($j\neq i$) is called an inter-loop. If a inter-loop passes through two charging stations $a,b\in \mathcal{C}$, the inter-loop with length $L$ is expressed as $r_{{\{a,b\}}}(L)$.
	The symbols used in this paper are listed in the table \ref{tab:CSC-notations}.

	\begin{table*}
		\label{tab:CSC-notations}
		\begin{centering}
			\caption{symbol table}
			\par\end{centering}
		\centering{}
		\begin{tabular}{ccl}
			\toprule
			Symbol & definition & \tabularnewline
			\midrule
			$N$ & Number of target points & \tabularnewline
			$M$ & number of sensors & \tabularnewline
			$K$ & number of charging stations & \tabularnewline
			$\mathcal{T}$ & target point collection $\{t_{1},t_{2},..,t_{N}\}$ & \tabularnewline
			$\mathcal{S}$ & sensor collection $\{s_{1},s_{2},..,s_{M}\}$ & \tabularnewline
			$\mathcal{C}$ & Charging station collection $\{c_{1},c_{2},..,c_{K}\}$& \tabularnewline
			$V$ & velocity of sensor & \tabularnewline
			$\omega(P)$ & the number of targets on path $P$ & \tabularnewline
			$T_c$ & Charge period of the sensor & \tabularnewline
			$T_t$ & sweep period of target point & \tabularnewline
			$L_c$ & $L_c=VT_c$ & \tabularnewline
			$L_t$ & $L_t=VT_t$ & \tabularnewline
			$SP(a,b)$ & the shortest path between charging stations $a$ and $b$ & \tabularnewline
			$d(a,b)$ & shortest path between point $a$ and point $b$ & \tabularnewline
			$\alpha$ & the approximation degree of the-rooted-orienteering algorithm & \tabularnewline
			$\beta$ & the approximation degree of the-rooted-team-orienteering algorithm & \tabularnewline
			$\gamma$ & s-t-orienteering algorithm approximation degree & \tabularnewline
			$\rho$ & approximation degree of mutual loop algorithm & \tabularnewline
			\bottomrule
		\end{tabular}
	\end{table*}

\section{Approximation algorithm for RCSC problem}
\label{sec:CSC-RCSC}

This section gives an approximation algorithm for the RCSC problem. For the RCSC problem, given an undirected complete graph $G(\mathcal{T}\cup\mathcal{C}, \mathcal{E})$, we need to find a loop for each sensor, and this loop satisfies constraints and makes $M$ sensors can  $T_t$-sweep  cover the largest total number of target points.

We use the integer programming formula \eqref{equ:CSC-limited-charger} to describe the problem. Among them, $y_t\in\{0,1\}$ indicates whether the target $t\in \mathcal{T}$ is covered; $r\subseteq \mathcal{R}$ is the path that meets the conditions; $x_i$ indicates whether to select the path $r_i$; $v_i$ is the number of sensors required for $T_t$-sweep covering the targets on path $r_i$. The constraint \eqref{equ:CSC-coverage-constraint} is the coverage constraint, and the constraint \eqref{equ:CSC-satisfication-constraint} is the feasibility constraint.

\begin{subequations}\label{equ:CSC-limited-charger}
	\begin{align}
		\text{max} & \sum_{t\in\mathcal{T}}y_{t} \nonumber\\
		\text{s.t.} & \sum_{i:t\in \mathcal{R}_i}x_{i}\ge y_{t} & \forall t\in \mathcal{T} \label{equ:CSC-coverage-constraint}\\
		& \sum_{r_i\in \mathcal{R}}v_{i}x_{i}\le M \label{equ:CSC-satisfication-constraint}\\
		& x_{i}\in\{0,1\} & \forall r_i\in \mathcal{R} \nonumber \\
		& y_{t}\in\{0,1\} & \forall t\in \mathcal{T} \nonumber
	\end{align}
\end{subequations}

Obviously, the number of candidate paths $r\in \mathcal{R}$ in Formula \eqref{equ:CSC-limited-charger} is exponent. However, if we can reduce the number of candidate paths in $\mathcal{R}$ to polynomials, the RCSC problem can be transformed into a maximum set coverage problem. 
In this maximum set coverage problem, a set of elements $\mathcal{T}={ t_1,t_2,...,t_n}$, a positive integer $M$, the set of sub-sets $\mathcal{R}=\{r_1,r_2,...,r_m\}$, in which $r_j\subset \mathcal{T}$  has its own cost $v_j$, are given, to find sub-sets with sum of whose cost not bigger than $M$ to cover the maximum number of elements.  The maximum set coverage problem is NP-hard\cite{gary1979computers}, but its polynomial-time approximation algorithm can be obtained.

\subsection{the RCSC algorithm when $T_c\ge T_t$}

\begin{figure*}[htb!]
	\centering
	\includegraphics[width=14 cm]{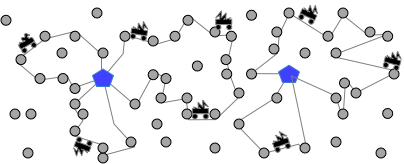}
	\caption{When $T_c\ge T_t$, under the RCSC algorithm, the sensor's sweeping path}
	\label{fig:RCSC-CT}
\end{figure*}

When $T_c\ge T_t$, $T_c/T_t=Q$ is a positive integer. Considering that the sensors travel periodically, the travel path $r\in \mathcal{R}$ of each sensor must be a loop. When the number of sensors  required for a path is the same, the longer the path is, the more targets it can cover. Therefore, the set of paths $\mathcal{R}$ can be reduced to include only paths whose length is an integer multiple of $L_t$.
Therefore, $r\in \mathcal{R}$ is loop $r_{c_i}( kL_t)$ which starts from any charging station $c_i\in\mathcal{C}$ and has a length of $k\times L_t$,  where $k=1,2,...,min\{Q,M\}$. That is, $\mathcal{R}=\{r_{c_i}(kL_t)\mid k\in[1,min\{Q,M\}],c_i\in\mathcal{C}\}$. And because the targets need to be $T_t$-sweep covered, the sensors need to be placed on loop $r_{c_i}(kL_t)$ at a distance of $VT_t$ to ensure the coverage period of the targets, so a loop $r_i$ with a length of $kL_t$ requires $v_i=k$ sensors. The sweeping path of a sensor is shown in figure \ref{fig:RCSC-CT}, the number of charging stations in the figure is 2, the number of sensors is 8, and $Q=3$.

Algorithm \ref{alg:CSC-RCSC-Lt} is implemented using a greedy strategy, gradually selecting the path $r\in \mathcal{R}$ with the most unity gain, and calling the sensors to cover it. Since the path is always found in the induced subgraph of uncovered targets, unity gain is the average number of targets covered by the sensors on the path. The steps of the algorithm are as follows:

\begin{enumerate}
	\item[1.] Let $\hat{\mathcal{T}}\subseteq\mathcal{T}$ be the set of targets that have not yet been covered, and construct a induced subgraph of graph $G$, $G'=(\hat{\mathcal{T}}\cup \mathcal{C},\mathcal{E}')$, where $\mathcal{E}'=\{(i,j)|i,j\in\hat{\mathcal{T}}\cup\mathcal{C},(i,j)\in\mathcal{E}\}$;
	\item[2.] Find the $c_i$ loop $r_{c_i}(kL_t)$ whose length is $kL_t$ and sensors on the loop cover the largetest number o targets in the graph $G'$, where $k=1,2, ...,min\{M',Q\}$, $M'$ is the number of sensors that have not yet been allocated;
	\item[3.] In the $K\times min\{M',Q\}$ loops $r_{c_i}(kL_t)$, to select the path $r_i$ of maximum unity gain. If the cost is $m_i$, distribute $m_i$ sensors on $r_i$ separated by $L_t$ to $T_t$-sweep cover targets on $r_i$;
	\item[4.] Repeat steps 1, 2, 3 until $M'=0$.
\end{enumerate}

From the \ref{subset:description} section, we know that the problem of finding the path with a limited length and the largest number of vertices is also called the orienteering problem, which is NP-hard. Algorithm \ref{alg:CSC-RCSC-Lt} calls the rooted-orienteering approximation algorithm  as a subroutine, with the approximation ratio is $\alpha$ ($0<\alpha<1$). The input is the induced subgraph $G'$ of the current uncovered targets and the the charging stations in the graph $G$, rooted node $c_i\in \mathcal{C}$ and the limited length $L$.  The final output of Algorithm \ref{alg:CSC-RCSC-Lt} is the set of covered target points $\mathcal{CT}$. We call the greedy algorithm that calls an approximation algorithm like Algorithm \ref{alg:CSC-RCSC-Lt} as the approximate greedy algorithm.

\begin{algorithm}[t]
	\SetAlgoNoLine \KwIn{ Figure $G(\mathcal{T}\cup\mathcal{C}, \mathcal{E})$, a set of sensors $\mathcal{S}$, its velocity is $V$, the target point The sweep period is $T_t$, and the sensor charging period is $T_c$. }\KwOut{Overriding target set $\mathcal{CT}$. }

	set $Q=\lfloor T_c/T_t\rfloor$;
	$\hat{\mathcal{T}}\leftarrow\mathcal{T}$\;
	$m_j \leftarrow0, \forall j\in[1,M]$\;
	$L_t=VT_t$, $L_c=VT_c$\;
	$\mathcal{CT}=\emptyset$\;
	set $r_j\leftarrow \emptyset, \forall j\in[1,M]$\;
	\For{ $j\leftarrow1$ to $M$ }{
		Construct an exported subgraph of graph $G$: $G'=(\hat{\mathcal{T}}\cup \mathcal{C},\mathcal{E}')$\;
		$Q'=min\{M-j+1,Q\}$\;
		\For{$i\leftarrow1$ to $K$}{
			\For {$k\leftarrow1$ to $Q'$}{
				Call the-rooted-orienteering($G',c_i,k\times L_t$) to find $c_i$-self-loop $r_(c_i)(kL_t)$\;
				$r_j\leftarrow$ $kL_t$ loop with the most number of target points covered by the unit sensor $r_{c_i}(kL_t)$, $m_j=k$\;
		}}

		Distribute the distance between the $m_j$ sensors by $L_t$ on $r_j$\;
		$j=j+m_j-1$\;
		$\hat{\mathcal{T}}\leftarrow \hat{\mathcal{T}}\setminus r_j$\;
		$\mathcal{CT}=\mathcal{CT}\cup (r_j\cap \mathcal{T})$\;}
	
	\Return $\mathcal{CT}$

	\caption{RCSC($T_c\ge T_t$)}
	\label{alg:CSC-RCSC-Lt}
\end{algorithm}

Suppose that the approximation solution obtained by this approximate greedy algorithm is $APP$, and $OPT$ is the optimal solution, the number of covered targets is $\omega(APP)$, $\omega(OPT)$. $r_i^{*}$ is the path with the most unity gain selected in the $i^{th}$ step of the greedy algorithm, and its cost is $v_i$. $r_i$ is the path selected in step $i$ of the approximation algorithm, with the same cost, i.e.,  $\omega(r_i)\ge\alpha\omega(r_i^{*})$. Here is a lemma \ref{lem:CSC-RCSC-greedy}:

\begin{lem}\label{lem:CSC-RCSC-greedy}
	In the approximate greedy algorithm, $\frac{1}{v_l}(\omega(\cup_{i=1}^{l}r_i)-\omega(\cup_{i=1}^{l-1}r_i) )\ge\frac{\alpha}{M}(\omega(OPT)-\omega(\cup_{i=1}^{l-1}r_i))$.
\end{lem}
\begin{proof}
	From the above approximate greedy steps, we can see that in the $l^{th}$ step, the $\alpha$-approximation algorithm, the-rooted-orienteering, is called, then
	\[\omega(\cup_{i=1}^{l}r_i)-\omega(\cup_{i=1}^{l-1}r_i)=\omega(r_l)\ge\alpha r_l^{ *};\]
	And because the set of targets with the weight of $\omega(OPT)-\omega(\cup_{i=1}^{l-1}r_i)$ can certainly be covered by $M$ sensors, and $r_l^{*}$ has the highest unity gain in the remaining candidate paths with the same cost $v_l$ as $r_l$.  According to the pigeonhole priciple,  $r_l^{*}$ satisfies
	$\omega(r_l^{*})/v_l\ge\frac{1}{M}(\omega(OPT)-\omega(\cup_{i=1}^{l-1}r_i))$, therefore,
	\[\frac{1}{v_l}(\omega(\cup_{i=1}^{l}r_i)-\omega(\cup_{i=1}^{l-1}r_i))\ge\frac{\alpha}{M}(\omega(OPT)-\omega(\cup_{i=1}^{l-1}r_i))\]
	established.
\end{proof}

The theorem \ref{thm:CSC-RCSC-greedy} can be easily derivated from the lemma \ref{lem:CSC-RCSC-greedy}.
\begin{thm}\label{thm:CSC-RCSC-greedy}
	The approximate greedy algorithm for the maximum set cover problem is an approximation algorithm with an approximate ratio of $(1-1/e)^{\alpha}$.
\end{thm}
\begin{proof}
	Assuming that the greedy algorithm has executed $m\le M$ interations, in the step $l$, $r_l$ ($l\le m$) is chosen, whose cost is $v_l$. $v_l$ is a positive integer and $v_l\le M -\sum_1^{l-1}v_{l-1}$.

	In the first step, $\omega(r_1)\ge\frac{\alpha v_1}{M}\omega(OPT)$ is established.
	
	Supposed $\beta_l=\frac{\alpha v_l}{M}$, according to the lemma \ref{lem:CSC-RCSC-greedy}, we know that, at step $l$,
	\begin{equation*}
		\begin{aligned}
			\omega(\cup_{i=1}^{l}r_i) &\ge\omega(\cup_{i=1}^{l-1}r_i)+\frac{\alpha v_l}{M}(\omega(OPT)-\omega(\cup_{i=1}^{l-1}r_i)) \\
			&\ge(1-\frac{\alpha v_l}{M})\omega(\cup_{i=1}^{l-1}r_i)+\frac{\alpha v_l}{M}\omega(OPT ) \\
			&\ge(1-\beta_l)\omega(\cup_{i=1}^{l-1}r_i)+\beta_l\omega(OPT) \\
			&\ge(1-\beta_l)(1-\beta_{l-1})\omega(\cup_{i=1}^{l-2}r_i) \\
			&\  \ +(\beta_l+\beta_{l-1}( 1-\beta_l)\omega(OPT) \\
			&\ge(1-\beta_l)(1-\beta_{l-1})\omega(\cup_{i=1}^{l-2}r_i)\\
			&\ \  +(1-(1-\beta_l)(1 -\beta_{l-1}))\omega(OPT) \\
			&\ge(1-\prod_{i\in[1,l]}(1-\beta_{i}))\omega(OPT)
		\end{aligned}
	\end{equation*}

	Then, after the $m$ step, $\sum_{i=1}^{m}v_i=M$, that is, $\sum_{i=1}^{m}\beta_i=\alpha$. therefore,
	\begin{equation*}
		\begin{aligned}
			\omega(APP) &=\omega(\cup_{i=1}^{m}r_i) \\
			&\ge(1-(\frac{1}{m}(\sum_{i\in[1,m]}(1-\beta_i)))^{m})\omega(OPT) \\
			&=(1-(1-\frac{\alpha}{m})^m)\omega(OPT) \\
			&>1-(1/e)^{\alpha}
		\end{aligned}
	\end{equation*}

	The first inequality sign comes from the average inequality, and the second inequality sign comes from $(1-\frac{\alpha}{m})^{\frac{m}{\alpha}}<1/e$. 
	
	The theorem can be proved.
\end{proof}

In each interation of Algorithm \ref{alg:CSC-RCSC-Lt}, it is needed to find the optimal path in the $K\times min\{Q,M'\}$ $c_i$-loops, then the-rooted-orienteering algorithm is called $K \times min\{Q,M'\}$ times. And  there is $M$ interations at most. The-rooted-orienteering algorithm is a polynomial approximation algorithm, and the time complexity is set to $O(L1)$, so the time complexity of Algorithm \ref{alg:CSC-RCSC-Lt} is $O(MK\times min\{ Q,M\}\times L1)$, it is polynomial-time solvable.

Since the best approximation ratio for the rooted orienteering problem in metric space is $\alpha=\frac{1}{2}$\cite{paul2020budgeted}, we get the theorem \ref{thm:CSC-RCSC-Lt -appr} as below.
\begin{thm}
	\label{thm:CSC-RCSC-Lt-appr}
	When $T_c\ge T_t$, the RCSC problem calls Algorithm \ref{alg:CSC-RCSC-Lt}, and the approximation ratio is $1-(1/e)^{\alpha}\approx \frac{2}{5}$.
\end{thm}

\subsection{the RCSC algorithm when $T_t> T_c$}

\begin{figure*}[htb!]
	\centering
	\includegraphics[width=14 cm]{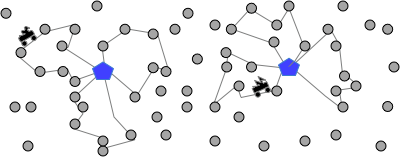}
	\caption{When $T_c< T_t$, under the RCSC algorithm, the sensor's sweeping path}
	\label{fig:RCSC-TC}
\end{figure*}

When $T_t>T_c$, $T_t/T_c=\hat{Q}$ is a positive integer. A sensor can start from one charging station $c_i\in\mathcal{C}$ and go through a set of $c_i$-loops with a length not exceeding $L_c$ and finally return to $c_i$ within a time period of $T_t$. The number of paths is exponential. Therefore, in Algorithm \ref{alg:CSC-RCSC-Lc}, only the path including $\hat{Q}$ loops starting from charging station $c_i\in\mathcal{C}$ and whose length is $L_c$ are considered, recorded as $r_{c_i}^{\hat{Q}}(Lc)$. Therefore,  $\mathcal{R}=\{r_{c_i}^{\hat{Q}}(Lc)|c_i\in\mathcal{C}\}$. The impression on the approximation ratio will be analysized later. The sweeping path of a sensor is shown in  Figure \ref{fig:RCSC-TC}, where the number of charging stations i is 2, the number of sensors is 2, and $\hat{Q}=3$.

Algorithm \ref{alg:CSC-RCSC-Lc} uses greedy method to gradually select the path  $r\in \mathcal{R}$ with the most unity gain, and call a sensor to cover. The  steps are as follows:

\begin{enumerate}
	\item[1.] Let $\hat{\mathcal{T}}\subseteq\mathcal{T}$ be the set of targets that have not yet been covered, and construct a induced subgraph of graph $G$, $G'=(\hat{\mathcal{T}}\cup \mathcal{C},\mathcal{E}')$, where $\mathcal{E}'=\{(i,j)|i,j\in\hat{\mathcal{T}}\cup\mathcal{C},(i,j)\in\mathcal{E}\}$;
	\item[2.] Find $\hat{Q}$ $c_i$-loops $r_{c_i}^{\hat{Q}}(L_c)$  with the largest number of targets in the graph $G'$. and deploy one sensor to cover;
	\item[3.] Repeat steps 1 and 2 until the number of sensors is 0.
\end{enumerate}

Finding the $r_{c_i}^{\hat{Q}}(Lc)$ that covers the most number of targets is  called the rooted team orienteering problem \cite{xu2020approximation}, which is NP-hard. therefore,
Algorithm \ref{alg:CSC-RCSC-Lc} calls the rooted-team-orienteering approximation algorithm for the rooted team orienteering problem as a subroutine, whose approximation ratio is $\beta$($0<\beta<1$). The input of Algorithm rooted-team-orienteering is the induced subgraph $G'$, the root  $c_i\in \mathcal{C}$, the path length $L$, and the number of paths $\hat{Q}$.

\begin{algorithm}[t]
	\SetAlgoNoLine \KwIn{ Figure $G(\mathcal{T}\cup\mathcal{C}, \mathcal{E})$, a set of sensors $\mathcal{S}$, its velocity is $V$, the target point The sweep period is $T_t$, and the sensor charging period is $T_c$. }\KwOut{Overriding target set $\mathcal{CT}$. }

	set $\hat{Q}=\lfloor T_t/T_c\rfloor$\;
	$\hat{\mathcal{T}}\leftarrow\mathcal{T}$\;
	$L_t=VT_t$, $L_c=VT_c$\;
	$\mathcal{CT}=\emptyset$\;
	set $r_j\leftarrow \emptyset, \forall j\in[1,M]$\;
	\For{ $j\leftarrow1$ to $M$ }{
		Construct an exported subgraph of graph $G$: $G'=(\hat{\mathcal{T}}\cup \mathcal{C},\mathcal{E}')$\;
		\For{$i\leftarrow1$ to $K$}{
			//The path consisting of $\hat{Q}$ loops with a length of $L_c$ and passing through $c_i$ that can cover the most target points\\
			Call the-rooted-orienteering($G',c_i,k\times L_t$) to find $c_i$-$\hat{Q}$ self-loop $r_{c_i}^{\hat{Q}}(L_c) $\;
			$r_j\leftarrow$ unit sensor covers the most number of target points $c_i$-$\hat{Q}$ self-loop$r_{c_i}^{\hat{Q}}(L_c)$\;}
		Use a sensor to cover the path formed by the $\hat{Q}$ circles of length $Lc$ $r_j$\;
		$\hat{\mathcal{T}}\leftarrow \hat{\mathcal{T}}\setminus r_j$\;
		$\mathcal{CT}=\mathcal{CT}\cup (r_j\cap \mathcal{T})$\;
	}

	\Return $\mathcal{CT}$

	\caption{RCSC($T_t\ge T_c$)}
	\label{alg:CSC-RCSC-Lc}
\end{algorithm}

Since the best approximation ratio of the rooted team orienteering problem in metric space is $\beta\approx0.394$\cite{xu2020approximation}, therefore, we get the theorem \ref{thm:CSC-RCSC-Lc-appr} as follows.
\begin{thm}
	\label{thm:CSC-RCSC-Lc-appr}
	When $T_t> T_c$, the RCSC problem calls Algorithm \ref{alg:CSC-RCSC-Lc}, and the approximation ratio is $1-(1/e)^{\hat{Q}\beta/(2\hat{Q }-1)}>\frac{1}{6}$.
\end{thm}

\begin{proof}
	On any charging station, two $c_i$ loops with the sum of whose length is less than $L_c$ can always be covered by a $c_i$-loop whose length is $L_c$. In the worst case, the optimal coverage path of a sensor is composed of $2\hat{Q}-1$ $c_i$-loops whose length is greater than $\frac{L_c}{2}$. Assume that the number of targets covered is $n$. Then according to the pigeonhole principle, $\hat{Q}$ loops $r_{c_i}(L_c)$ can always be found  to make the number of targets covered greater than $\frac{n}{2}$. Assuming that the number of targets covered by the optimal $r_{c_i}^{\hat{Q}}(L_c)$ in the current induced subgraph $G'$ is $n_L$, then $n_L\ge \frac{\hat{ Q}}{2\hat{Q}-1}n$. Because the subroutine the-rooted-team-orienteering is a $\beta$-approximation algorithm, the number of targets covered by the selected path must be not less than $\frac{\hat{Q}\beta}{2\hat{Q} -1}n$. According to the theorem \ref{thm:CSC-RCSC-greedy}, the approximation ratio of Algorithm \ref{alg:CSC-RCSC-Lc} is $1-(1/e)^{\hat{Q}\beta/( 2\hat{Q}-1)}$.

	In Algorithm \ref{alg:CSC-RCSC-Lc}, each interation needs to find the optimal solution in the $K$ paths, i.e., the-rooted-team-orienteering algorithm is called $K$ times in each interation,  $M$ interation in total. Set the time complexity of Algorithm the-rooted-team-orienteering  to $O(L2)$, then the time complexity of Algorithm \ref{alg:CSC-RCSC-Lc} is $O(KM\times L2)$, polynomial-time solvable.

	Therefore, Algorithm \ref{alg:CSC-RCSC-Lc} is an approximation algorithm, and the approximation ratio is related to $\hat{Q}$. When $\hat{Q}=1$, the approximation ratio is about $1/3$, When $\hat{Q}\rightarrow
	\infty$, the approximation ratio is approximately $\frac{1}{6}$.
\end{proof}

\section{the approximation algorithm for 2-CSC problem}
\label{sec:CSC-2CSC}

When the sensor can go to any charging station for charging,  a variety of combinations make paths many, which greatly increases the complexity of the chargeable sweep coverage problem. This article analyzes one of the simpler cases: the 2-chargeable sweep coverage problem when the number of charging stations is two.

Suppose $\mathcal{C}=\{a,b\}$, the shortest path $SP(a,b)$ length is $d(a,b)\le L_c$, otherwise, there is no mutual loop in the candidate paths $\mathcal{R}$. Similarly, the 2-CSC problem is expressed by the formula \eqref{equ:CSC-limited-charger} and considered as a maximum set coverage problem. At this time, the candidate paths $\mathcal{R}$ include self-loops $r_a$ and $r_b $ and mutual loops $r_{\{a,b\}}$. The algorithms are given for  the 2-CSC problem in two cases when $T_c\ge T_t$ and when $T_t>T_c$ respectively.

\subsection{the algorithm for 2-CSC problem when $T_c\ge T_t$}

\begin{figure*}[htb!]
		\centering
		\includegraphics[width=14 cm]{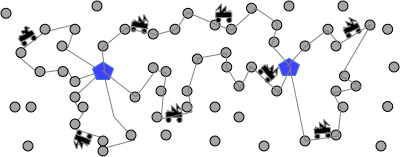}
		\caption{When $T_c\ge T_t$, under the 2CSC algorithm, the sensor's sweeping path}
		\label{fig:2CSC-CT}
	\end{figure*}

When $T_c\ge T_t$, $T_c/T_t=Q$ is a positive integer. Similar to the RCSC problem, the travelling paths of a sensor  $r\in \mathcal{R}$ must be a loop. When the number of sensors $v_i$ required for the path is the same, the longer the path, the more the number of targets can be covered. Therefore, the set of paths $\mathcal{R}$ can be reduced to only include paths with the length $k\times L_t$. Supposed $\bar{Q}=\lceil\frac{d(a,b)}{L_t}\rceil$, the path set includes self-loop $\{r_a(i\times L_t), r_b(i\times L_t ) \mid i\in[1,min\{Q,M\}]\}$, mutual-loop $\{ r_{\{a,b\}}(j\times L_t) \mid j\in[2 \bar{Q},min\{M,2Q\}]\}$. When $M<2\bar{Q}$, the mutual-loop does not exist. The cost of self-loop $r_c(i\times L_t)$ is $i$, $c\in\{a,b\}$; the cost of mutual-loop $r_{\{a,b\}}(j\times L_t)$ is $j$. Suppose $j_1+j_2=j$, if $j_1$ and $j_2$ are not integers, then $j_1$ and $j_2$ can have countless combinations. Therefore, Algorithm \ref{alg:CSC-2CSC-Lt} only considers the mutual loops composed of two paths with length that are integer multiples of $L_t$, denoted as $r_{\{a,b\}}((j_1+j_2)\times L_t)$. The set of candidate paths is $\mathcal{R}=\{r_a(i\times L_t),r_b(i\times L_t) \mid i\in[1,min\{Q,M\}]\}\cup \{r_{\{a,b\}}((j_1+j_2)\times L_t)\mid j_1,j_2\in Z^{+},j_1+j_2=j\in[2\bar{Q},min\{M,2Q\}]\}$. The sweeping path of the sensor is shown in Figure \ref{fig:2CSC-CT}, the number of charging stations in the figure is 2, the number of sensors is 8, and $Q=2$.
	
	Algorithm \ref{alg:CSC-2CSC-Lt} uses a greedy strategy to gradually select the path $r\in \mathcal{R}$ with the most unity gain, and call the sensor to cover it. The specific steps of the algorithm are as follows:
	
	\begin{enumerate}
	\item[1.] Let $\hat{\mathcal{T}}\subseteq\mathcal{T}$ be the set of targets that have not yet been covered, and construct a induced subgraph of graph $G$, $G'=(\hat{\mathcal{T}}\cup \mathcal{C},\mathcal{E}')$, where $\mathcal{E}'=\{(i,j)|i,j\in\hat{\mathcal{T}}\cup\mathcal{C},(i,j)\in\mathcal{E}\}$;
	\item[2.] In the graph $G'$, find an mutual loop $r_{\{a,b\}}((j_1+j_2)L_t)$ with the average largest number of targets covered by the sensors, where $j_1,j_2\in Z^{+}$, $j_1+j_2=j$, $j=2\bar{Q},,...,min\{M',2Q \}$, $M'$ is the number of sensors that have not yet been allocated;
	\item[3.] In the graph $G'$, find the $c_i$ self loop $r_{c_i}(kL_t)$ whose length is $kL_t$ and cover the largetest average number of targets, where $c_i\in\{a,b\}$$k=1,2, ...,min\{M',Q\}$, $M'$ is the number of sensors that have not yet been allocated;
	\item[4.] Compare the unity gain of the path obtained in steps 2 and 3, choose the path with the larger unity gain, if the cost is $m_i$, then call $m_i$ sensors to distribute on $r_i$ eparated by $L_t$, $T_t$-sweep coverage of the targets on $r_i$;
	\item[5.] Repeat steps 1-4 until $M'=0$.
\end{enumerate}

In the step 2, it is NP-hard to find the mutual-loop $r_{\{a,b\}}((j_1+j_2)L_t)$ whose length is $(j_1+j_2)L_t$ and covers the largest number of targets. We have known a s-t orienteering problem, aiming to find a path with the length not greater than $L$ and coveloop the most number of targets, is an NP-hard problem. Its approximation algorithm, s-t-orienteering, whose  input is the current induced subgraph $G'$ , two root nodes $\{a,b\}$ and limited length $L$. Assuming that the approximation ratio of the approximation algorithm is $\gamma$, we design an approximation algorithm to find the mutual loop $r_{\{a,b\}}((j_1+j_2)L_t)$ covering the largest number of targets ( Algorithm \ref{alg:CSC-s-t-loop}).

\begin{algorithm}[htb!]
\SetAlgoNoLine \KwIn{ Figure $G'(\hat{\mathcal{T}}'\cup\mathcal{C}, \mathcal{E'})$, two $st$ path lengths $L_1$, $L_2 $}\KwOut{a $st$ mutual loop$\{P_1;P_2\}$. }
$P_1=s-t-orienteering((G',s,t,L_1))$\;
Construct an exported subgraph of $G'$:$G''=(\{\hat{\mathcal{T'}}\setminus P_1\}\cup \mathcal{C},\mathcal{E}'')$ , Where $\mathcal{E}''=\{(i,j)|i,j\in V(G''),(i,j)\in\mathcal{E}'\}$;\;
$P_2=s-t-orienteering((G'',s,t,L_2))$\;
\Return $\{P_1;P_2\}$
\caption{s-t-loop}
\label{alg:CSC-s-t-loop}
\end{algorithm}

This is also a greedy approximation algorithm. From the  proof of the theorem \ref{thm:CSC-RCSC-greedy}: $\omega(APP)\ge (1-(1-\frac{\gamma}{m}) ^m)\omega(OPT)$, when $m=2$, the approximation ratio of Algorithm \ref{alg:CSC-s-t-loop} is $1-(1-\frac{\gamma}{2})^2$. The lemma \ref{lem:CSC-huhuanlu} is as follows:
\begin{lem}
\label{lem:CSC-huhuanlu}
Supposed the approximation ratio of the s-t-orienteering algorithm is $\gamma$, then under the greedy approximation algorithm, the approximation ratio of Algorithm \ref{alg:CSC-s-t-loop} is $\rho=1-(1- \frac{\gamma}{2})^2$.
\end{lem}

\begin{algorithm}[htb!]
\SetAlgoNoLine \KwIn{ Figure $G(\mathcal{T}\cup\mathcal{C}, \mathcal{E})$, a set of sensors $\mathcal{S}$, its velocity is $V$, the target point The sweep period is $T_t$, and the sensor charging period is $T_c$. }\KwOut{Overriding target set $\mathcal{CT}$. }

set $Q=T_c/T_t$\;
$\hat{\mathcal{T}}\leftarrow\mathcal{T}$\; $m_j \leftarrow0, \forall j\in[1,M]$ \;
set $r_j\leftarrow \emptyset, \forall j\in[1,M]$\;
$\bar{Q}=\lceil \frac{d(a,b)}{L_T}\rceil$\;

\For{ $j\leftarrow1$ to $M$ }{
	Construct an exported subgraph of graph $G$: $G'=(\hat{\mathcal{T}}\cup \mathcal{C},\mathcal{E}')$\;
	\For {$k_1\leftarrow\bar{Q}$ to min $\{M-j+1-\bar{Q}, Q\}$}{
		\For {$k_2\leftarrow\bar{Q}$ to min $\{M-j+1-k_1, Q\}$}{
			Call Algorithm \ref{alg:CSC-s-t-loop} to find the $a-b$ mutual loop $r_{\{a,b\}}(k_1L_t\oplus k_2L_t)$ on the graph $G'$\;
			$tr_1\leftarrow$ the mutual loop with the largest unity gain, $v_1=k_1+k_2$\;
	}}
	\For {$k\leftarrow1$ to min $\{M-j+1,Q\}$}{
		Call the-rooted-orienteering($G',a,k\times L_t$) to find $a$-self-loop $r_(a)(kL_t)$\;
		Call the-rooted-orienteering($G',b,k\times L_t$) to find $b$-self-loop $r_(b)(kL_t)$\;
		Compare the two, $tr_2\leftarrow$ unity gain maximum self-loop, $v_2=k$\;}
	$i=argmax_{i\in\{1,2\}}\omega(tr_i)/m_i$, $m_j=v_i$, $r_j=tr_i$\;
	$j=j+m_j-1$\;
	$\hat{\mathcal{T}}\leftarrow \hat{\mathcal{T}}\setminus r_j$\;
	Distribute the distance between the $m_j$ sensors by $L_t$ on $r_j$\;
	$\mathcal{CT}=\mathcal{CT}\cup (r_j\cap \mathcal{T})$\;}
\Return $\mathcal{CT}$

\caption{2-CSC($T_c\ge T_t$)}
\label{alg:CSC-2CSC-Lt}
\end{algorithm}

Supposed that the number of targets covered by Algorithm \ref{alg:CSC-2CSC-Lt} is $\omega(APP)$, and the number of targets covered by the optimal solution is $\omega(OPT)$. $r_i^{*}$ is the path with the most unity gain selected in the $i^{th}$ interation, whose cost is $v_i$. $r_i$ is the path chosen by Algorithm \ref{alg:CSC-2CSC-Lt} in the  $i^{th}$ interation with the same cost. From the lemma \ref{lem:CSC-RCSC-greedy}, we know that if the path chosen for the $l$th time is a self-loop, 
\[\frac{1}{v_l}(\omega(\cup_{i=1}^{l}r_i)-\omega(\cup_{i=1}^{l-1}r_i))\ge\frac{\alpha}{M}(\omega(OPT)-\omega(\cup_{i=1}^{l-1}r_i))\]
where $\alpha$ is the approximate ratio of Algorithm the-rooted-orienteering.
In Algorithm \ref{alg:CSC-2CSC-Lt}, if the path chosen for the $l^{th}$ interation is a mutual loop, the lemma \ref{lem:CSC-2CSC-greedy} is below.
\begin{lem}
	\label{lem:CSC-2CSC-greedy}
	In Algorithm \ref{alg:CSC-2CSC-Lt}, if the path $r_l$ selected in the $l$ iteration is a mutual loop, where $v_l$ is the cost of $r_l$,  \[\frac{1}{v_l-1}(\omega(\cup_{i=1}^{l}r_i)-\omega(\cup_{i=1}^{l-1}r_i))\ge\frac{\rho}{M}(\omega(OPT)-\omega(\cup_{i=1}^{l-1}r_i))\]
\end{lem}
\begin{proof}
	Since the optimal path is always found on the uncovered target set $\mathcal{T}$ and charging station $\mathcal{C}$, \[\omega(r_l)=\omega(\cup_{i=1 }^{l}r_i)-\omega(\cup_{i=1}^{l-1}r_i)\]. 
	Assuming that $r_l^{*}$ is the optimal path currently selected, 
	and it consists of two $a-b$ paths that are not integer multiples of $L_t$,  let the cost of $r_l^{*}$ be $k$. From the lemma \ref{lem:CSC-RCSC-greedy},
	$r_l^{*}$ satisfies $\omega(r_l^{*})/k\ge\frac{1}{M}(\omega(OPT)-\omega(\cup_{i=1}^{l -1}r_i)$.
	In Algorithm \ref{alg:CSC-2CSC-Lt}, find a mutual loop $r_{l2}^{*}$ composed of the two $a-b$ paths with integer multiples of length $L_t$, and set the lengths to be $k_1L_t$, $k_2L_t$ respectively.
	Then $v_l=k_1+k_2=k+1$. Both $a-b$ paths are longer than the two $a-b$ paths of the original $r_l^{*}$, so $\omega(r_{l2}^{*})\ge \omega(r_l^{*}) $. According to  Lemma \ref{lem:CSC-huhuanlu}, $\omega(r_l)\ge \rho\omega(r_{l2}^{*})$. Thus, 
	\begin{equation*}
		\begin{aligned}
			\frac{1}{v_l-1}(\omega(\cup_{i=1}^{l}r_i)-\omega(\cup_{i=1}^{l-1}r_i)) &= \frac{1}{k}\omega(r_l)\\
			&\ge \frac{\rho}{k}\omega(r_{l2}^{*}) \\
			&\ge \frac{\rho}{M}(\omega(OPT)-\omega(\cup_{i=1}^{l-1}r_i))
		\end{aligned}
	\end{equation*}
	
	The theorem is provable.
	
\end{proof}

\begin{thm}\label{thm:CSC-2CSC-greedy}
	Algorithm \ref{alg:CSC-2CSC-Lt} is an approximation algorithm with an approximation ratio of approximately $\frac{1}{4}$.
\end{thm}
\begin{proof}
	Assuming that the greedy algorithm has been executed  $m\le M$ interations, $r_l$ ($l\le m$) is chosen in the  $l^{th}$ interation.  whose cost $v_l$ is a positive integer and $v_l\le M -\sum_1^{l-1}v_{l-1}$.

	If $r_l$ is not a mutual loop, it is known from the lemma \ref{lem:CSC-RCSC-greedy} that \[\frac{1}{v_l}(\omega(\cup_{i=1}^{l}r_i )-\omega(\cup_{i=1}^{l-1}r_i))\ge\frac{\alpha}{M}(\omega(OPT)-\omega(\cup_{i=1}^ {l-1}r_i))\]
	If $r_l$ is a mutual loop, it can be obtained by the lemma \ref{lem:CSC-2CSC-greedy},
	\[\frac{1}{v_l-1}(\omega(\cup_{i=1}^{l}r_i)-\omega(\cup_{i=1}^{l-1}r_i))\ge\frac{\rho}{M}(\omega(OPT)-\omega(\cup_{i=1}^{l-1}r_i))\]

	Then, when $r_l$ is not a mutual loop, set $\beta_l=\frac{\alpha v_l}{M}$; when $r_l$ is a mutual loop, set $\beta_l=\frac{\alpha (v_l-1 )}{M}$.
	After $m$ iterations, $\sum_{i=1}^{m}v_i=M$, that is, $\sum_{i=1}^{m}\beta_i=(1-\frac{m'}{M})\alpha$, where $m'\le m$ is the number of mutual loops. Only when the number of mutual loops is $m'\le M/2\bar{Q}$, and  $\bar{Q}\ge2$, there will be a mutual loop composed by two $a-b$ paths with a length that is not an integer multiple of $L_t$. Therefore, $\frac{m'}{M}\le\frac{1}{4}$. According to Theorem \ref{thm:CSC-RCSC-greedy},
	\begin{equation*}
		\begin{aligned}
			\omega(APP) &=\omega(\cup_{i=1}^{m}r_i) \\
			&\ge(1-(\frac{1}{m}(\sum_{i\in[1,m]}(1-\beta_i)))^{m})\omega(OPT) \\
			&=(1-(1-\frac{min\{\alpha,\rho\}}{m}(1-\frac{m'}{M}))^m)\omega(OPT) \\
			&\ge (1-(1/e)^{(1-\frac{m'}{M})\times min\{\alpha,\rho\}})\omega(OPT) \\
			&\ge (1-(1/e)^{\frac{3}{4}\times min\{\alpha,\rho\}})\omega(OPT)
		\end{aligned}
	\end{equation*}

	Choose the $1/(2+\epsilon)$-approximation algorithm in the paper \cite{chekuri2012improved} as Algorithm s-t-orienteering\footnote{At present, there have been a variety of approximation algorithms for the st-orienteering problem, and the approximation ratios are 1/4\cite{blum2007approximation}, 1/3\cite{bansal2004approximation}, $1/(2+\epsilon)$\cite{chekuri2012improved}. }, set $\epsilon=0.5$, then $\gamma=0.4$. According to Lemma \ref{lem:CSC-huhuanlu}, $\rho=0.36$. $\alpha=\frac{1}{2}$\cite{paul2020budgeted}. Then $\omega(APP)\ge \frac{1}{4}\omega(OPT)$.
	The time complexity of Algorithm \ref{alg:CSC-2CSC-Lt} is $O(M\times min\{Q^2,M^2\}\times max\{O(L1),O(L2)\})$, where $O(L1)$ is the time complexity of Algorithm the-rooted-orienteering, and $O(L2)$ is the time complexity of Algorithm  the-rooted-team-orienteering. Therefore Algorithm \ref{alg:CSC-2CSC-Lt} is a polynomial time algorithm. The theorem can be proved.
\end{proof}

\subsection{the algorithm for  2-CSC when $T_t> T_c$}

\begin{figure*}[htb!]
	\centering
	\includegraphics[width=14 cm]{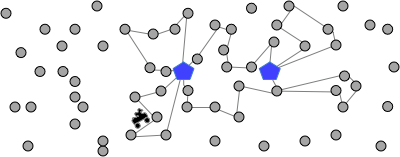}
	\caption{When $T_c< T_t$, under 2CSC algorithm, the sensor's sweeping path}
	\label{fig:2CSC-TC}
\end{figure*}

When $T_t>T_c$, $T_t/T_c=\hat{Q}$ is a positive integer. A sensor can start from any charging station $c\in\{a,b\}$, go through a set of paths no longer than $L_c$, and finally return to $c$ within a time period of $T_t$. There are countless such loops, so in the algorithm \ref{alg:CSC-2CSC-Lc}, only the loops composed of $\hat{Q}$ paths with a length of $L_c$ is considered, recorded as $r_{\{a,b\}}(\hat{Q}\odot L_c)$, i.e., $\mathcal{R}=\{r_{\{a,b\}}(\hat{Q}\odot L_c)\}$, which is shown in figure \ref{fig:2CSC-TC}, where the number of charging stations in the figure is 2, the number of sensors is 1, and $\hat{Q}=6$.

As shown in Figure \ref{fig:2CSC-TC}, the reasonable travel path of a single sensor satisfies one of the following situations:
\begin{enumerate}
	\item[1.] it is only composed of $a$ self-loop;
	\item[2.] it is only composed of $b$ self-loops;
	\item[3.] it is composed of $a$ self-loop and $a-b$ mutual loop;
	\item[4.] it is composed of $b$ self-loop and $a-b$ mutual loop;
	\item[5.] it consists of three kinds of loops: $a$ self-loop, $b$ self-loop, and $a-b$ mutual loop.
\end{enumerate}

For $M$ sensors, there can be $M\hat{Q}$ paths with a length of $L_c$. From the reasonable path of a single sensor described above, the reasonable composition of the $M\hat{Q}$ paths satisfies one of situations below:

\begin{enumerate}
	\item[1.] There is at least one mutual loop $r_{\{a,b\}}$;
	\item[2.] Without a mutual loop, it is composed of $k_1$ $a$ self-loop and $k_2$ $b$ self-loop, where $k_1+k_2=M\hat{Q}$, and $k_1/ \hat{Q}$ and $k_2/\hat{Q}$ are integers;
\end{enumerate}

The randomly selected $M\hat{Q}$ paths may constitute an unreasonable compostion of paths, which  consists of $k_1$ $a$ self-loops and $k_2$ $b$ self-loops, where $k_1+k_2= M\hat{Q}$, $k_1/\hat{Q}$ or $k_2/\hat{Q}$ are not integers. Therefore, steps 7-8 of Algorithm \ref{alg:CSC-2CSC-Lc} are used to avoid the unreasonable compositon of paths. The  steps of Algorithm \ref{alg:CSC-2CSC-Lc} are as follows:

\begin{enumerate}
	\item[1.] Let $\hat{\mathcal{T}}\subseteq\mathcal{T}$ be the set of targets that have not yet been covered, and construct an induced subgraph of graph $G$, $G'=(\hat{\mathcal{T}}\cup \mathcal{C},\mathcal{E}')$, where $\mathcal{E}'=\{(i,j)|i,j\in\hat{\mathcal{T}}\cup\mathcal{C},(i,j)\in\mathcal{E} \}$;
	\item[2.] If the number of chosen paths is not more than $M\hat{Q}-2$, find a mutual loops $r_{\{a,b\}}(2L_c)$ composed by two paths with length of $L_c$ and covering the largest number of targets in the graph $G$;
	\item[3.] Find the $a$ self-loop $r_a(L_c)$ with a length of $L_c$ and covering the largest number of targets in the graph $G'$;
	\item[4.] Find the $b$ self-loop $r_b(L_c)$ with a length of $L_c$ and covering the largest number of targets in the graph $G'$;
	\item[5.] Compare the paths obtained in steps 2, 3, and 4 to choose the path with the largest unity gain;
	\item[6.] Repeat steps 1-5 until you find $M\hat{Q}-2$ paths;
	\item[7.] If mutual loop $r_{\{a,b\}}(L_c)$ exist, repeat steps 1-5 until you find $M\hat{Q}$ paths, and go to step 9.
	\item[8.] If mutual loop $r_{\{a,b\}}(L_c)$ does not exist, set the current number of $a$ self-loops to $k_1$ and the number of  $b$ self-loops to $k_2$,
	\begin{enumerate}
		\item[a.] If $k_1/\hat{Q}$ and $k_2/\hat{Q}$ are both integers,
		\begin{enumerate}
			\item[i.] When $\hat{Q}=2$, after performing step 1, find the mutual loop $r_{\{a,b\}}$ with the largest gain, or two $a$ self-loops , or two $b$ self-loops, and select  two paths with the maximum gain.
			\item[ii.] When $\hat{Q}>2$, perform steps 1 and 2 to find the mutual loop with the maximum gain.
		\end{enumerate}
		\item[b.] If $k_1/\hat{Q}$ is an integer and $k_2/\hat{Q}$ is not an integer, repeat steps 1, 2, 4, and select the path with the largest unity gain until you find $M\hat{Q}$ paths,
		\item[c.] If $k_2/\hat{Q}$ is an integer and $k_1/\hat{Q}$ is not an integer, repeat steps 1, 2, and 3, and select the path with maximum unity gain until it finds $M\hat{Q}$ paths,
		\item[d.] If both $k_2/\hat{Q}$ and $k_1/\hat{Q}$ are not integers, repeat steps 1, 3, 4, and 8, and select the path of maximum unity gain until Find $M\hat{Q}$ paths,
	\end{enumerate}
	\item[9.] After finding $M\hat{Q}$ paths, assign sensors.
	\begin{itemize}
		\item Let $k_1$ be the number of $a$ self-loops in $\mathcal{SR}$; $k_2$ is the number of $b$ self-loops in $\mathcal{SR}$; $k_3$ is the number of mutual loops $r_{\{a,b\}}$ in $\mathcal{SR}$; 
		\item allocate $\lfloor k_1/\hat{Q}\rfloor$ sensors  to cover $\lfloor k_1/\hat{Q}\rfloor\times \hat{Q}$ $a$ self-loops;
		\item allocate $\lfloor k_2/\hat{Q}\rfloor$ sensors to cover $\lfloor k_2/\hat{Q}\rfloor\times \hat{Q}$ $b$ self-loops;
		\item allocate $\lfloor 2k_3/\hat{Q}\rfloor$ sensors on the $\{a,b\}$ mutual-loops, covering $\lfloor k_3/\hat{Q}\rfloor\times \hat{ Q}$$\{a,b\}$ mutual loops;
		\item The remaining path can be covered by the remaining 1 or 2 sensors.
	\end{itemize}
\end{enumerate}

\begin{algorithm*}[htb!]
	\SetAlgoNoLine \KwIn{ Figure $G(\mathcal{T}\cup\mathcal{C}, \mathcal{E})$, a set of sensors $\mathcal{S}$, its velocity is $V$, the target point The sweep period is $T_t$, and the sensor charging period is $T_c$. }\KwOut{path collection $\mathcal{SR}$. }

	set $\hat{Q}=T_t/T_c$; $\hat{\mathcal{T}}=\mathcal{T}$; $L_c=VT_c$; $L_t=VT_t$\;
	\For{$i\leftarrow 1$ to $M$}{
		\For{$j\leftarrow 1$ to $\hat{Q}$}{
			Construct an induced subgraph of graph $G$: $G'=(\hat{\mathcal{T}}\cup \mathcal{C},\mathcal{E}')$\;
			\eIf{$(i-1)\times \hat{Q}+(j-1)==M\hat{Q}-2$ and there is no mutual loop in $\mathcal{SR}$}{
				Let $k_1$ be the current number of $\mathcal{SR}$ self-loop $r_a$; $k_2$ is the current number of $\mathcal{SR}$ self-loop $r_b$\;
				\uIf{$k_1 \mod \hat{Q}== 0$ and $k_2 \mod \hat{Q}== 0$ }{
					\eIf{$\hat{Q}==2$}{
						Find the maximum unity gain mutual loop $r_{\{a,b\}}(2L_c)$\;
						Find two maximum gains $a$-self-loop $r1_a(L_c)$ and $r2_a(L_c)$\;
						Find two maximum gains $a$-self-loop $r1_b(L_c)$ and $r2_b(L_c)$\;
						Compare the above, choose the best gain, and put the path into $\mathcal{SR}$\;}{
						$r\leftarrow$Maximum unity gain mutual loop$r_{\{a,b\}}(2L_c)$,$\mathcal{SR}=\mathcal{SR}\cup \{r\}$\;}
					}
				\uElseIf{$k_1 \mod \hat{Q}== 0$ and $k_2 \mod \hat{Q}\neq 0$ }{
					Find the maximum unity gain mutual loop $r_{\{a,b\}}(2L_c)$\;
					Find two maximum gains $b$-self-loop $r1_b(L_c)$ and $r2_b(L_c)$\;
					Compare the above, choose the best gain, and put the path into $\mathcal{SR}$\;}
				\uElseIf{$k_1 \mod \hat{Q}\neq 0$ and $k_2 \mod \hat{Q}== 0$}{
					Find the maximum unity gain mutual loop $r_{\{a,b\}}(2L_c)$\;
					Find two maximum gains $a$-self-loop $r1_a(L_c)$ and $r2_a(L_c)$\;
					Compare the above, choose the best gain, and put the path into $\mathcal{SR}$\;}
				\uElse(\tcc*[f]{$k_1 \mod \hat{Q}\neq 0$ and $k_2 \mod \hat{Q}\neq 0$}){
					First find the maximum gain $a$-self-loop $r1_a(L_c)$, then find the maximum gain $b$-self-loop $r2_b(L_c)$\;
					First find the maximum gain $b$-self-loop $r1_b(L_c)$, then find the maximum gain $a$-self-loop $r2_a(L_c)$\;
					Compare the above, choose the best gain, and put the path into $\mathcal{SR}$\;}
				}{
				Call the algorithm \ref{alg:CSC-s-t-loop} to find the maximum unity gain mutual loop $r_{\{a,b\}}(2L_c)$\;
				Call the-rooted-orienteering to find the maximum gain self-loop $r_a(Lc)$\;
				Call the-rooted-orienteering to find the maximum gain self-loop $r_b(Lc)$\;
				Compare the above, select the path with the largest unity gain $r$\;
				$\mathcal{SR}=\mathcal{SR}\cup\{r\}$\;
				\Return $\mathcal{SR}$\;
			}

			$\hat{\mathcal{T}}\leftarrow \hat{\mathcal{T}}\setminus r$\; 
	}
}

	Let $k_1$ be the number of $a$ self-loops in $\mathcal{SR}$; $k_2$ is the number of $b$ self-loops in $\mathcal{SR}$; $k_3$ is the number of mutual loops $r_{\{a,b\}}$ in $\mathcal{SR}$\; 
	Allocate $\lfloor k_1/\hat{Q}\rfloor$ sensors on the $a$ self-loop to cover $\lfloor k_1/\hat{Q}\rfloor\times \hat{Q}$ $a$ self-loop Loop\;
	Allocate $\lfloor k_2/\hat{Q}\rfloor$ sensors on the $b$ self-loop, covering $\lfloor k_2/\hat{Q}\rfloor\times \hat{Q}$ $b$ self-loop Loop\;
	Allocate $\lfloor 2k_3/\hat{Q}\rfloor$ sensors on the $\{a,b\}$ mutual loop to cover $\lfloor k_3/\hat{Q}\rfloor\times \hat{Q} $a$\{a,b\}$ mutual loops\;
	The remaining path can be covered by the remaining 1 or 2 sensors\;
	
	\Return $\mathcal{SR}$

	\caption{2-CSC($T_t> T_c$)}
	\label{alg:CSC-2CSC-Lc}
	
\end{algorithm*}

\begin{thm}
	Algorithm \ref{alg:CSC-2CSC-Lc} is $1-(1/ e)^{\hat{Q}\times min\{\alpha,\rho\}/(2\hat{Q}-1)}$-approximation algorithm.
\end{thm}

\begin{proof}
	Two $c_i$ self-loops whose length is less than $L_c$ can always be covered by a $c_i$-self-loop with length $L_c$, and an mutual-loop formed by two paths less than $L_c/2$ It can be covered with a self-loop of length $L_c$. Therefore, in the worst case, the optimal sweeping path of a sensor is composed of $2\hat{Q}-1$ paths longer than $\frac{L_c}{2}$, and satisfy one of reasonable situations. Assume that the number of targets covered by the optimal deployment is $n$. Then according to the pigeonhole principle, $\hat{Q}$ loops can always be found in it to make the number of targets more than $\frac{n}{2}$, and the loops satisfy one of situations of sensors' combination paths (analysis omitted). Assuming that the maximum number of targets covered by $\hat{Q}$ paths with a length of $L_c$ is $n_L$, then $n_L\ge \frac{\hat{Q}}{2\hat{Q}- 1}n$. Because Algorithm the-rooted-orienteering is a $\alpha$-approximation algorithm, and the approximation ratio of Algorithm s-t-loop  is $\rho$, the number of targets covered by the selected paths must be greater than or equal to $\frac{\hat{Q }\times min\{\alpha, \rho\}}{2\hat{Q}-1}n$. According to  Theorem \ref{thm:CSC-RCSC-greedy}, the approximation ratio of Algorithm \ref{alg:CSC-RCSC-Lc} is $1-(1/e)^{\hat{Q}\times min\{\alpha,\rho\}/(2\hat{Q}-1)}$.
	\end{proof}
	
	If $\alpha=\frac{1}{2}$, $\gamma=\frac{2}{5}$, then $\rho=0.36$, Theorem \ref{thm:CSC-2CSC-Lc-appr} is obtained.
	\begin{thm}
		\label{thm:CSC-2CSC-Lc-appr}
		When $T_t> T_c$, the 2-CSC problem calls Agorithm \ref{alg:CSC-2CSC-Lc}, the approximation ratio of which is related to $\hat{Q}$. When $\hat{Q}=1$, the approximation ratio is approximately $0.3$; when $\hat{Q}$=2, the approximation ratio is 0.2134; when $\hat{Q}\rightarrow
		\infty$, the approximation ratio is approximately $\frac{1}{6}$.
	\end{thm}

\section{Conclusion}
\label{sec:CSC-conclution}

This paper presents the model of  the sweep coverage problem of rechargeable sensors for the first time. Through the comparison of existing problems, the difficulty of the CSC problem  is discussed. Besides, two types of chargeable sweep coverage problems under different constraints are proposed: RCSC problem and 2-CSC problems,  both of which are NP-hard problems. Approximation algorithms are provided for them. The RCSC problem and the 2-CSC problem are modeled as the maximum coverage problems, by reducing the candidate paths to polynomials and calling the existing approximation algorithms as subroutines, we obtain the approximation algorithms. The future work is to consider other algorithmic techniques, improve the approximation ratio of these two problems, and further study the general chargeable sweep coverage problem. The sweep coverage problem of rechargeable sensors has rich research content:
\begin{enumerate}
	\item The general chargeable sweep coverage problem. For example, the number of charging stations is greater than 2 and the sensors can go to any charging stations for charging, how to schedule the sensors to cover the maximum number of targets;
	\item The placement  of the charging stations. For example, how to place the charging stations so that the given sensors can cover the maximum number of targets;
	\item The minimum sensor coverage problem: The locations of the charging stations and the targets are known, how to schedule the sensors so that the minimum number of sensors can sweep cover all targets while satisfying the charging demand.
	\item If battery capacities of given sensors are inconsistent, how can the chargeable sweep coverage problem be modeled and solved?
\end{enumerate}

\section*{Acknowledgment}

...

\bibliographystyle{plain}
\bibliography{Chargeable_Sweep_Coverage_Problem}

\begin{thebibliography}{10}

\bibitem{bansal2004approximation}
Nikhil Bansal, Avrim Blum, Shuchi Chawla, and Adam Meyerson.
\newblock Approximation algorithms for deadline-tsp and vehicle routing with
  time-windows.
\newblock In {\em Proceedings of the thirty-sixth annual ACM symposium on
  Theory of computing}, pages 166--174, 2004.

\bibitem{blum2007approximation}
Avrim Blum, Shuchi Chawla, David~R. Karger, Terran Lane, Adam Meyerson, and
  Maria Minkoff.
\newblock Approximation algorithms for orienteering and discounted-reward tsp.
\newblock {\em SIAM Journal on Computing}, 37(2):653--670, 2007.

\bibitem{chekuri2012improved}
Chandra Chekuri, Nitish Korula, and Martin Pal.
\newblock Improved algorithms for orienteering and related problems.
\newblock {\em ACM Transactions on Algorithms}, 8(3):1--27, 2012.

\bibitem{chen2006orienteering}
Ke~Chen and Sariel Har-Peled.
\newblock The orienteering problem in the plane revisited.
\newblock In {\em Proceedings of the twenty-second annual symposium on
  Computational geometry}, pages 247--254, 2006.

\bibitem{chen2016efficient}
Zhiyin Chen, Xudong Zhu, Xiaofeng Gao, Fan Wu, Jian Gu, and Guihai Chen.
\newblock Efficient scheduling strategies for mobile sensors in sweep coverage
  problem.
\newblock In {\em Sensing, Communication, and Networking (SECON), 2016 13th
  Annual IEEE International Conference on}, pages 1--4. IEEE, 2016.

\bibitem{cheng2008sweep}
Weifang Cheng, Mo~Li, Kebin Liu, Yunhao Liu, Xiangyang Li, and Xiangke Liao.
\newblock Sweep coverage with mobile sensors.
\newblock In {\em Parallel and Distributed Processing, 2008. IPDPS 2008. IEEE
  International Symposium on}, pages 1--9. IEEE, 2008.

\bibitem{collins2013optimal}
Andrew Collins, Jurek Czyzowicz, Leszek Gasieniec, Adrian Kosowski, Evangelos
  Kranakis, Danny Krizanc, Russell Martin, and Oscar Morales~Ponce.
\newblock Optimal patrolling of fragmented boundaries.
\newblock In {\em Proceedings of the twenty-fifth annual ACM symposium on
  Parallelism in algorithms and architectures}, pages 241--250. ACM, 2013.

\bibitem{gao2018approximation}
X.~F. Gao, J.~H. Fan, F.~Wu, and G.~H. Chen.
\newblock Approximation algorithms for sweep coverage problem with multiple
  mobile sensors.
\newblock {\em Ieee-Acm Transactions on Networking}, 26(2):990--1003, 2018.

\bibitem{gao2020cooperative}
Xiaofeng Gao, Jiahao Fan, Fan Wu, and Guihai Chen.
\newblock Cooperative sweep coverage problem with mobile sensors.
\newblock {\em IEEE Transactions on Mobile Computing}, pages 1--1, 2020.

\bibitem{gary1979computers}
Michael~R Gary and David~S Johnson.
\newblock Computers and intractability: A guide to the theory of
  np-completeness, 1979.

\bibitem{gorain2016solving}
B.~Gorain and P.~S. Mandal.
\newblock Solving energy issues for sweep coverage in wireless sensor networks.
\newblock {\em Discrete Applied Mathematics}, 228:130--139, 2017.

\bibitem{gorain2019approximation}
B.~Gorain and P.~S. Mandal.
\newblock Approximation algorithms for barrier sweep coverage.
\newblock {\em International Journal of Foundations of Computer Science},
  30(3):425--448, 2019.

\bibitem{gorain2014approximation}
Barun Gorain and Partha~Sarathi Mandal.
\newblock Approximation algorithms for sweep coverage in wireless sensor
  networks.
\newblock {\em Journal of Parallel and Distributed Computing},
  74(8):2699--2707, 2014.

\bibitem{gorain2014line}
Barun Gorain and Partha~Sarathi Mandal.
\newblock Line sweep coverage in wireless sensor networks.
\newblock In {\em 2014 Sixth International Conference on Communication Systems
  and Networks (COMSNETS)}, pages 1--6. IEEE, 2014.

\bibitem{gorain2015approximation}
Barun Gorain and Partha~Sarathi Mandal.
\newblock Approximation algorithm for sweep coverage on graph.
\newblock {\em Information Processing Letters}, 115(9):712--718, 2015.

\bibitem{huang2018efficient}
Peihuang Huang, Wenxing Zhu, Kewen Liao, Timos Sellis, Zhiyong Yu, and Longkun
  Guo.
\newblock Efficient algorithms for flexible sweep coverage in crowdsensing.
\newblock {\em IEEE Access}, 6:50055--50065, 2018.

\bibitem{laporte1984two}
Gilbert Laporte, Martin Desrochers, and Yves Nobert.
\newblock Two exact algorithms for the distance-constrained vehicle routing
  problem.
\newblock {\em Networks}, 14(1):161--172, 1984.

\bibitem{nagarajan2012approximation}
V.~Nagarajan and R.~Ravi.
\newblock Approximation algorithms for distance constrained vehicle routing
  problems.
\newblock {\em Networks}, 59(2):209--214, 2012.

\bibitem{pasqualetti2012cooperative}
Fabio Pasqualetti, Antonio Franchi, and Francesco Bullo.
\newblock On cooperative patrolling: Optimal trajectories, complexity analysis,
  and approximation algorithms.
\newblock {\em IEEE Transactions on Robotics}, 28(3):592--606, 2012.

\bibitem{paul2020budgeted}
Alice Paul, Daniel Freund, Aaron Ferber, David~B Shmoys, and David~P
  Williamson.
\newblock Budgeted prize-collecting traveling salesman and minimum spanning
  tree problems.
\newblock {\em Mathematics of Operations Research}, 45(2):576--590, 2020.

\bibitem{shi2016adaptive}
Tuo Shi, Siyao Cheng, Zhipeng Cai, and Jianzhong Li.
\newblock Adaptive connected dominating set discovering algorithm in
  energy-harvest sensor networks.
\newblock In {\em IEEE INFOCOM 2016-The 35th Annual IEEE International
  Conference on Computer Communications}, pages 1--9. IEEE, 2016.

\bibitem{shi2017constructing}
Tuo Shi, Siyao Cheng, Jianzhong Li, and Zhipeng Cai.
\newblock Constructing connected dominating sets in battery-free networks.
\newblock In {\em IEEE INFOCOM 2017-IEEE Conference on Computer
  Communications}, pages 1--9. IEEE, 2017.

\bibitem{shi2018coverage}
Tuo Shi, Jianzhong Li, Hong Gao, and Zhipeng Cai.
\newblock Coverage in battery-free wireless sensor networks.
\newblock In {\em IEEE INFOCOM 2018-IEEE Conference on Computer
  Communications}, pages 108--116. IEEE, 2018.

\bibitem{wang2016hybrid}
Cong Wang, Ji~Li, Yuanyuan Yang, and Fan Ye.
\newblock A hybrid framework combining solar energy harvesting and wireless
  charging for wireless sensor networks.
\newblock In {\em IEEE INFOCOM 2016-The 35th Annual IEEE International
  Conference on Computer Communications}, pages 1--9. IEEE, 2016.

\bibitem{wu2019task}
L.~G. Wu, Y.~H. Xiong, M.~Wu, Y.~He, and J.~H. She.
\newblock A task assignment method for sweep coverage optimization based on
  crowdsensing.
\newblock {\em Ieee Internet of Things Journal}, 6(6):10686--10699, 2019.

\bibitem{xu2020approximation}
Wenzheng Xu, Zichuan Xu, Jian Peng, Weifa Liang, Tang Liu, Xiaohua Jia, and
  Sajal~K Das.
\newblock Approximation algorithms for the team orienteering problem.
\newblock In {\em IEEE INFOCOM 2020-IEEE Conference on Computer
  Communications}, pages 1389--1398. IEEE, 2020.

\bibitem{yu2017participant}
Zhiyong Yu, Jie Zhou, Wenzhong Guo, Longkun Guo, and Zhiwen Yu.
\newblock Participant selection for t-sweep k-coverage crowd sensing tasks.
\newblock {\em World Wide Web}, 21(3):741--758, 2017.

\bibitem{zhang2019timely}
D.~Zhang, D.~Zhao, and H.~D. Ma.
\newblock On timely sweep coverage with multiple mobile nodes.
\newblock {\em 2019 Ieee Wireless Communications and Networking Conference
  (Wcnc)}, 2019.

\bibitem{zhao2012mobile}
Dong Zhao, Huadong Ma, and Liang Liu.
\newblock Mobile sensor scheduling for timely sweep coverage.
\newblock In {\em 2012 IEEE Wireless Communications and Networking Conference
  (WCNC)}, pages 1771--1776. IEEE, 2012.

\end{thebibliography}

\end{document}